\documentclass[12pt]{article}
\usepackage{graphicx}
\usepackage{listings}
\usepackage{mathtools}
\usepackage{tikz-cd} 

\usepackage{url}
\begin{document}
%

%\title{Contribution Title\thanks{Supported by organization x.}}
\title{Reflective Metagraph Rewriting \\  as a Foundation\\  for an AGI "Language of Thought" \\
-- \\
{\it \small Toward a Formalization of OpenCog Hyperon's MeTTa Language \\
in Terms of Algebraic Metagraph Rewriting} \\
}

%\titlerunning{Neural Embedding of Semantic Hypergraphs}

\author{Ben Goertzel}

\date{%
    SingularityNET Foundation \\%
    \today
}
%
%\authorrunning{Goertzel et al.}
% d \institute{SingularityNET Foundation }
\maketitle              % typeset the header of the contribution
\begin{abstract}
MeTTa (Meta Type Talk) is a novel programming language created for use in the OpenCog Hyperon AGI system.   It is designed as a meta-language with very basic and general facilities for handling symbols, groundings, variables, types, substitutions and pattern matching.   Primitives exist for creating new type systems and associated DSLs, the intention being that most MeTTa programming takes place within such DSLs.   

 Informally, MeTTa is Hyperon's lowest-level "language of thought" -- the meta-language in which algorithms for learning more particular knowledge representations, will operate, and in which these algorithms themselves may be represented.  Tractable representation of a variety of knowledge and cognitive process types in this sort of formalism has been explored in a long history of publications and software systems.

Here we explain how one might go about formalizing the MeTTa language as a system of metagraph rewrite rules, an approach that fits in naturally to the Hyperon framework given that the latter's core component is a distributed metagraph knowledge store (the Atomspace).   The metagraph rewrite rules constituting MeTTa programs can also be represented as metagraphs, giving a natural model for MeTTa reflection and self-modifying code.   

Considering MeTTa programs that compute equivalences between execution traces of other MeTTa programs allows us to model spaces of MeTTa execution traces using Homotopy Type Theory.   Considering the limit of MeTTa programs mapping between execution traces of MeTTa programs that map between execution traces of MeTTa programs that $\ldots$, we find that a given MeTTa codebase is effectively modeled as an $\infty-\textrm{groupoid}$, and the space of all MeTTa codebases as an $(\infty, 1)-\textrm{topos}$.  This topos is basically the same as the so-called "Ruliad" previously derived from rewrite rules on hypergraphs, in a discrete physics context.

The formalization of MeTTA as metagraph rewrite rules may also provide a useful framework for structuring the implementation of efficient methods for pattern matching and equality inference within the MeTTa interpreter.
\end{abstract}
%%%%%%%%%%%%%%%%%%%%%%%%%%%%%%%%%%%%%%%%%%%%%%%%%%%%%%%5
%%%%%%%%%%%%%%%%%%%%%%%%%%%%%%%%%%%%%%%%%%%%%%%%%%%%%%%5

\tableofcontents

\section{Introduction}

OpenCog is an integrative software framework aimed to foster the creation of systems with AGI at the human level and beyond \cite{EGI1}  \cite{EGI2} \cite{Hyperon}.   The core design feature is a distributed metagraph called the Atomspace, consisting of nodes and (hyper)links labeled with various types and other information.   Multiple AI algorithms originating from different AI paradigms (logical reasoning, probabilistic programming, attractor neural networks, evolutionary learning) are then implemented leveraging Atomspace for internal representation as well as cross-algorithm communication.   An associated cognitive systems theory explains how this approach can encompass all the central aspects of human intelligence, with a focus on the "cognitive synergy" among the different algorithms and different modes of memory organization involved.

Concurrently with leveraging the existing "OpenCog Classic" codebase to carry out various practical functions, a portion of the OpenCog community is now engaged in developing a new version of the system called OpenCog Hyperon  \cite{Hyperon}.   The motives underlying Hyperon are multiple, including desires for greater scalability and usability, and to leverage tools and methods from branches of math such as dependent type theory and intuitionistic and paraconsistent logic at deeper levels in the system design.

A key ingredient of the Hyperon design is the MeTTa (Meta Type Talk) programming language, which serves as the key structural principle of the Hyperon version of the Atomspace, as a language for internal use by Hyperon algorithms, and (with appropriate surface syntax) as an end-user language for developers to use in coding algorithms and applications for Hyperon.   Informally, MeTTa is Hyperon's lowest-level "language of thought" -- the meta-language in which algorithms for learning more particular knowledge representations, will operate, and in which these algorithms themselves may be represented.   

Why we consider MeTTa a viable "language of thought" for human-like (and some forms of non-human-like) AGI will not be richly explicated here; however, tractable representation of a variety of knowledge and cognitive process types in various precursors to the MeTTa formalism (e.g. OpenCog Classic's Atomese) has been explored in a long history of publications and software systems.   The recent papers \cite{GTGI}  \cite{goertzel2021patterns}  \cite{goertzel2020paraconsistent} building on prior work such as \cite{EGI1} \cite{EGI2} tell this part of the story.   The goal here is to add a necessary piece to this overall story via going beyond "MeTTa as a language for manipulating knowledge metagraphs" to "MeTTa as knowledge represented in knowledge metagraphs, enabling knowledge metagraphs to flexibly self-transform in a cognitively useful and meaningful way."   This latter perspective is by no means conceptually surprising in the context of the overall OpenCog and Hyperon programmes, but nevertheless requires its own detailed explication.

The initial MeTTa specification \cite{MeTTaSpec} is semi-formalized and articulates the language mainly by way of examples accompanied with textual descriptions.   Implementation of a prototype MeTTa interpreter is underway.    The end-user syntax for MeTTa is not yet decided; the prototype and early documentation uses a vaguely LISP-like syntax but this is fairly likely not to survive, especially as there is currently considerable discussion of design directions such as integration of MeTTa into Python and Julia as a framework, and use of a graphical interactive expression editor.

As part of the development and analysis of MeTTa and the overall Hyperon system, it will be useful to have a more formal presentation of the MeTTa language.   This article does not provide such a formal presentation in detail, but it outlines what we believe is the most natural approach to creating one, via partially fleshing out a framework presenting MeTTa as a system of metagraph rewrite rules.    We will describe in fair depth one route to creating such a formalization, and then briefly explore some of the follow-on consequences expected to follow from this formalization, including connections to topos theory and strategies for efficient implementation of core aspects of a MeTTa interpreter.

Conceptually the MeTTa language as described in \cite{MeTTaSpec} could be formulated and implemented in contexts other than metagraph rewriting.   However, the actual initial  practical intention is to implement MeTTa in the context of the Hyperon design, which has to do with storing MeTTa scripts in the Hyperon Atomspace (metagraph), and using these scripts to act on said Atomspace etc.  So for near and medium term purposes, a formalization in terms of metagraph rewrite rules seems like a broad enough scope.

We reiterate that the aim of this sort of formalization is pragmatic as well as theoretical.   To implement MeTTa as a programming language in practice, one has to some mapping of MeTTa into underlying operations to be carried out in the computer; and mediating this mapping via mapping MeTTa into a relatively simple syntactic formalism like metagraph rewrite rules has promise to make the process more orderly and comprehensible.

\section{MeTTa Language Overview}

This section gives a brief overview of some key aspects of the MeTTA language and its core library; for a fuller treatment see \cite{MeTTaSpec}.

The textual MeTTa syntax used here, drawn from the MeTTa spec, is merely preliminary and illustrative, and will potentially be changed radically before the language hits any users.

\subsection{MeTTa Vocabulary}

MeTTA symbols come in three forms: grounded symbols, constant ungrounded symbols, and variables

Each grounded symbol is associated with some software object which is able to identify if a given digital entity fits the ''grounding criterion'' of that symbol.   Some grounded symbols are provided in a core library, e.g. numbers and strings; others may be added by the developer.

A grounded symbol $F$ may be a grounded function symbol, meaning that evaluating $(F, A)$ according to $F$'s grounding involves feeding $A$ into the code providing this grounding.   The code grounding $F$ may either ask for the value obtained after feeding $A$ to the MeTTa interpreter, or it may ''quote'' $A$ and ask for $A$ as a literal expression (an object of ''meta-language type Expression'').

Expressions are formed by combining symbols, using the concatenation operator that lets us group symbols $a$and $b$ into the list $(a b)$.   Concatenation can be recursively nested into constructs like $((a,b),c)$, which makes expressions into familiar-looking trees of symbols.  

The use of MeTTa in practice is centered on a few key grounded symbols provided in the core library:

\begin{itemize}
\item The $:$ type-assignment operator and $\textrm{Type}$ symbol for specifying types
\item The $->$ arrow operator for specifying type restrictions
\item The $=$ ''equality'' operator for specifying expression-evaluation relationships
\end{itemize}

Atomspaces are sets of expressions.  The default way of storing Atomspaces is as metagraphs (which involves deduplication of repeated sub-expressions among multiple expressions in the same Atomspace).   An alternative would be to store a set of expressions as a forest of dags, which then requires some other way of accounting for duplicated sub-dags.

\subsection{Types}

A MeTTa Type is basically ''a symbol whose associated constraints, as regards the usage of symbols and expressions that belong to it, get proactively checked by the MeTTA interpreter in the context of evaluation.''   If a symbol is not a Type, then checking any constraints regarding its usage has to be done via explicitly invoked and scheduled inference.   If a symbol is a Type, then checking of its constraints is done automatically along with the evaluation of expressions involving the symbol.  This is a ''control'' issue more than an ''underlying logic'' issue.

Saying e.g.

{\tt\begin{small}\begin{lstlisting}
( : Human Type)
\end{lstlisting}\end{small}}

\noindent uses the keyword Type to specify that Human should be interpreted as a MeTTa type.  Saying e.g.

{\tt\begin{small}\begin{lstlisting}
(: Bob Human) 
\end{lstlisting}\end{small}}

\noindent then specifies that the symbol Bob is of type Human.  Saying

{\tt\begin{small}\begin{lstlisting}
(: a (-> B A))
\end{lstlisting}\end{small}}

\noindent specifies that if $b$ has type $B$ then $(a b)$ has type $A$ (or if $b$ is undefined in type then $(a,b)$ is undefined in type).

\subsection{Further Enrichments}

Other enrichments aside from types as typically construed may be treated similarly.  An example would be vector embeddings.   A user could define an enrichment of type ''EmbeddingVector'' and associate it with symbols as follows:

{\tt\begin{small}\begin{lstlisting}
(: EmbeddingVector Enrichment)
(: KPCAVector EmbeddingVector)
(: BobV KPCAVector)
(Association (Bob BobV))
\end{lstlisting}\end{small}}

There will certainly be implementation subtleties associated with such enrichments -- e.g. properly handling the changes to the grounding of ''match'' needed to make matching properly handle additional enrichments like this (because unlike most symbol groundings, grounding of ''match'' is core to the operation of the MeTTa interpreter).

\subsection{Pattern Matching}

A ''pattern'' is an expression which is specified for use as a query in the context of one or more Atomspaces.   

Matching a pattern $P$ against an Atomspace or other expression-set $A$

{\tt\begin{small}\begin{lstlisting}
(match P A)
\end{lstlisting}\end{small}}

\noindent returns a set of all the expressions in $A$ that can be made to instantiate $P$ by assigning particular values to the variables in $P$.

Pattern matching automatically invokes type checking; so if some variables in $P$ are associated with types, they will only be assigned matches that are compatible with their types.

When $P$ and some expression $E$ in $A$ both contain variables, the matching process must try to bind both the variables in $P$ and those in $E$; but precedence is given to binding the variables in $P$.

Patterns can be combined via various operators.   The role of disjunction or conjunction operators is clear.   To what extent negation will play a role, and in what form (e.g. subject to what form of guarding) is a current research topic.

The matching function

{\tt\begin{small}\begin{lstlisting}
((transform P T) A)
\end{lstlisting}\end{small}}

\noindent involves an ''output template'' $T$ that is an expression containing a subset of the variables in $P$.  The transform process performs matching of $P$ against $A$, and then returns a version of $T$ with variables assigned to their corresponding values from $P$.  So e.g.

{\tt\begin{small}\begin{lstlisting}
(transform ((has Sam $o) $o) ( ? (has Sam balloon ) (has Sam ball)))
\end{lstlisting}\end{small}}

\noindent returns the list (balloon ball).

Intermediate expressions created during the matching or transformation process are by default put into an auxiliary Atomspace, separate from the Atomspace $A$ being matched against.

The recursive matching algorithm behind the ''match'' function may involve various subtleties, depending on the type systems according to which ''match'' needs to perform checking.   E.g. some complex unification problems will in many cases appear behind the scenes here.

''Match'' and ''Transform'' may be considered as grounded symbols, grounded by the underlying algorithms.

Variations of basic matching might be introduced to handle various sorts of enrichments, e.g.

{\tt\begin{small}\begin{lstlisting}
(matchEV P A e)
\end{lstlisting}\end{small}}

\noindent for some numerical threshold e, might perform matching using embedding vectors, accepting only matches for which the proximity of the query and the match in embedding space exceeds $e$.

\subsection{Equalities}

The MeTTA equality operator 

{\tt\begin{small}\begin{lstlisting}
(= A B)
\end{lstlisting}\end{small}}

\noindent has the rough semantics ''$A$ can be evaluated as $B$.''   It is not symmetric.

E.g. a Boolean random variable may be configured via

{\tt\begin{small}\begin{lstlisting}
(= bin 0)
(= bin 1)  
\end{lstlisting}\end{small}}

A simple deterministic function might be set up as

{\tt\begin{small}\begin{lstlisting}
(= (double $x) ($x $x))
\end{lstlisting}\end{small}}

Equality should be considered as a built-in Type.   However, the static type checking carried out automatically during type checking does not incorporate inference using equalities (including type equalities).   (If it did, this would basically turn pattern matching into full-on MeTTa interpretation.  Instead, in the MeTTa design, the equality inference is put into the interpreter loop, which invokes static pattern matching.)

\subsection{Interpretation}

The MeTTa interpretation process is triggered by providing the interpreter with a query, which is either:

\begin{itemize}
\item A pattern matching request
\item A request for execution of a grounded symbol (on certain arguments)
\item A request for evaluation of an expression (including an ungrounded symbol)
\end{itemize}

Any of these options may recursively trigger additional steps drawn from any of the options.

In the third case, a request to evaluate expression $E$ will be interpreted as an equality query of the form

{\tt\begin{small}\begin{lstlisting}
(= E $t)
\end{lstlisting}\end{small}}

\noindent (whose interpretation may then involve multiple recursive steps).

For the interpretation process to effectively evaluate the truth of expressions, the additional equality

{\tt\begin{small}\begin{lstlisting}
( = (: $t Type) (transform (: $w $t) True))
\end{lstlisting}\end{small}}

\noindent is needed to instantiate the mapping from inhabited types to true expressions (i.e. if the Type is nonempty then it should be evaluated to True).

\section{Graph and Metagraph Rewrite Rules}

Existing algebraic formalizations of graph rewriting can be extended to handle rewriting on directed labeled metagraphs; the only significant step needed is to define an appropriate notion of homomorphism on the latter, which reflects the semantics of types and enrichments used as labels and also deals appropriately with the complexities of combining ordered and unordered edge targets.

\subsection{Algebraic Formalization of Graph Rewriting}

We begin with the single-pushout (SPO) formalization of graph rewriting, as outlined and utilized e.g. in  \cite{GrGen} 

Paraphrasing the above paper: A SPO graph rewrite rule $p : L \xrightarrow{r} R$ consists of a pattern graph $L$, a replacement graph $R$ and a partial graph homomorphism $r$ between $L$ and $R$. An application of $p$ to a host graph $G$ is called a direct derivation. It requires a partial graph homomorphism $m$ from $L$ to $G$ , which is called a match. The direct derivation leads to a result graph $H$ .  For each node or edge $x$ in $L$ there exists a corresponding node or edge in $G$, namely $m(x)$.   The overall process is depicted in the following diagram:

\[ \begin{tikzcd}
L \arrow{r}{\varphi} \arrow[swap]{d}{m} & R \arrow{d}{\mu} \\%
G \arrow{r}{\rho}& H
\end{tikzcd}
\]

Note that the mapping m does not need to be injective.   The preservation morphism $r$ determines what happens to $m(x$): It maps all items from $L$ to $R$, which are to remain in $G$ during the application of the rule. The images under $m$ of all items in $L$ which have no image under $r$ are to be deleted. The others are retained. Items in $R$ which have no pre-image under $r$ are added to $H$ . Note that in general the morphism $\rho$  from $G$ to $H$ is neither surjective nor total. It is partial, because nodes from $G$ may be deleted to get $H$. Homomorphism $\rho$ can be non-surjective, because new nodes may be introduced in H' -- these new nodes are not in the image of $\rho$ but in the image of $\mu$.

The  main alternative approach is the double-pushout approach (DPO)  \cite{DPO} .  The main difference is how dangling links are dealt with when a node is deleted.   DPO  only deletes a node when the rule specifies the deletion of all adjacent edges as well, whereas SPO just disposes the adjacent edges automatically.

In the DPO approach, roughly speaking, a production is given by $p = (L, K, R)$, where $L$ and $R$ are the left- and right-hand side graphs and $K$ is the common interface of $L$ and $R$, i.e. their intersection. The left-hand side
$L$ represents the preconditions of the rule, while the right-hand side $R$ describes the postconditions. $K$ describes a graph part which has to exist to apply the rule, but which is not changed. $L/K$ describes the part which is to be deleted, and $R/K$ describes the part to be created.   A direct graph transformation with a production $p$ is defined by first finding a match m of the left-hand side $L$ in the current host graph $G$ such that $m$ is structure-preserving.

When a direct graph transformation with a production $p$ and a match m is performed, all the vertices and edges which are matched by $L/K$ are removed from $G$. The removed part is not a graph, in general, but the remaining structure $D := (G/m(L)) \cup m(K)$ still has to be a legal graph, i.e., no edges should be left dangling. This means that the match m has to satisfy a suitable gluing condition, which makes sure that the gluing of $L/K$ and $D$ is equal to $G$.   In the second step, the graph $D$ is glued together with $R/K$ to obtain the derived graph $H$.   Since $L$ and $R$ can overlap in $K$, the submatch occurs in the original graph $G$ and is not deleted in the first step, i.e. it also occurs in the intermediate graph $D$. For gluing newly created vertices and edges into $D$, the graph $K$ is used. This defines the gluing items at which $R$ is inserted into $D$.

\[ \begin{tikzcd}
L \arrow[leftarrow]{r} \arrow[swap]{d}{m} & K \arrow{r} \arrow{d} & R \arrow{d}\\%
G \arrow[leftarrow]{r} & D \arrow{r} & H
\end{tikzcd}
\]

\subsection{Algebraic Formalization of Directed Labeled Metagraph Rewriting}

SPO and DPO are about graphs, but the generalization to hypergraphs or metagraphs (including directed labeled metagraphs) is immediate.  One merely needs an appropriate notion of homomorphism.   

A hypergraph homomorphism is a map from the vertex set of one hypergraph to another such that each edge maps to one other edge.   Metagraphs are similar but slightly tricker.  

We consider here {\it labeled} metagraphs, by which we mean metagraphs whose edges are potentially labeled with {\it types} and also other enrichments, such as e.g. numerical vectors or tensors, where:

\begin{itemize}
\item The types may be drawn from any appropriate type theory
\begin{itemize}
\item e.g. in a Hyperon context we are working with variations on probabilistic dependent type theory, inspired by \cite{goertzel2020paraconsistent} and various references therein, and the Idris 2 type system  \cite{brady2020idris} and other sources
\item Key is that there is a notion of type inheritance so that with two types $T$ and $T_1$ we can evaluate whether the inheritance relation $T<T_1$ holds.  
\end{itemize}
\item For other enrichments used as part of labels, it will be convenient if they come with a notion of homomorphism \begin{itemize}
\item For instance, if tensors are used as enrichments, then linear transformations on tensor space are homomorphisms
\item If there is no more sophisticated notion of homomorphism available for a particular sort of enrichment, then the identity mapping can serve as homomorphism operation, meaning that metagraph homomorphisms will need to preserve these enrichments precisely
\end{itemize}
\end{itemize}

\noindent Note that, fundamentally, the types and enrichments involved in a metagraph may be considered as shorthands for pointers to metagraphs embodying the information in these types/enrichments.  In the end one could make do without labels via encoding label/enrichment information in connectivity structure, but this would be awkward and not a useful approach for practical implementation in current hardware/software infrastructures.

A (directed labeled) metagraph homomorphism would then be naturally defined as  follows.   Suppose the edges of the metagraphs of concern are labeled with types $T$ drawn from type-system $\mathcal{T}$ and additional enrichments $S$ drawn from the algebra $\mathcal{S}$ on which a notion of homomorphism is defined.    A metagraph homomorphism $\Phi$ is then a map from the edge set of one metagraph $M$ to edge set of another metagraph $M'$ with the property that there exists some homomorphism $\Psi$ on $\mathcal{S}$ so that

\begin{itemize}
\item If  in  $M$
\begin{itemize}
\item $E$ links to $F$ with type $T$ and enrichment $S$
\end{itemize}
\noindent then in $M'$
\begin{itemize}
\item  $\Phi(E)$ links to $\Phi(F)$ with type $T'$, enrichment $S'$
\end{itemize}
\noindent where the inheritance $T'<T$ holds, and $S ' = \Psi(S)$
\item  If in $M$
\begin{itemize}
\item $E$ links to $F$ with index $i_1$ and  type $T_1$
\item $E_1$ links to $F_1$ with index $i_2>i_1$ and type $T_2$
\end{itemize}
\noindent, then in $M'$ ,  
\begin{itemize}
\item $\Phi(E)$ links to $\Phi(F)$ with index $j_1$ and type $T'_1$, and enrichment $S'_1$
\item $\Phi(E_1)$ links to $\Phi(F_1)$ with index $j_2>j_1$ and type $T'_2$ and enrichment $S'_2$

\end{itemize}
\noindent where $T'_1<T_1$ , $T'_2<T_2$, $S_1' = \Psi(S_1)$, $S_2' = \Psi(S_2)$.
\end{itemize}

\noindent Note that "Edges" in this case include vertices as degenerate edges.

With this definition of a homomorphism, the formalization of algebraic graph rewriting given above, in both the SPO and DPO scenarios, carries over perfectly to the labeled typed metagraph case.

Basically, this definition of a labeled directed metagraph homomorphism provides a definition for what it means for one sub-metagraph to form a "pattern" that "matches" another sub-metagraph.   This is a more basic form of pattern-matching than the pattern-matching in the MeTTa language, which explicitly involves variable binding, but it's a suitable under-layer both for MeTTa pattern-matching and other aspects of MeTTa.

The treatment of enrichments beyond types implicit in this definition is mildly subtle: basically $M$ is said to match $M'$ if there is a match in all aspects besides enrichments, and if there is a homomorphism that maps the enrichments of $M$ into the corresponding enrichments of $M'$.   This requirement is necessary and sufficient to ensure that the commutation diagrams presented above for SPO and DPO rewriting will still work in the context of enriched metagraph edges.

\section{MeTTa as Metagraph Rewriting}

We now review the basic concepts needed to represent the MeTTa language as a collection of metagraph rewrite rules.   Full formalization along these lines will be left for a subsequent, auxiliary document.

At the high level, MeTTa may be understood in metagraph terms via

\begin{itemize}
\item Portraying MeTTa expressions as dags (similarly to the standard portrayal e.g. of LISP expressions as trees)
\item Portraying the MeTTa grounding domains as directed labeled metagraphs, where the ''labels'' may include types and also potentially other enrichments such as numerical vectors or tensors 
\end{itemize}

\noindent Obviously this represents taking the first step toward metagraph-rewrite-rule representation of MeTTa via observing: If the expressions and the grounding domains are all metagraphs (a dag being a sub-case of a metagraph), then one can conceive and analyze MeTTA entirely in terms of metagraph rewriting.
 
Interpretation of a MeTTa expression is then a matter of a MeTTa expression dag (which is part of a metagraph) being interpreted as a metagraph rewrite rule, and enacted on the metagraph which it specifies as its target.  This may include reflexive self-rewriting.

So from a metagraph view, the crux of MeTTa can then be viewed as: A way of interpreting certain sub-metagraphs of a directed labeled metagraph $M$, as rewrite rules to be enacted on some sub-metagraph of $M$.   What MeTTa provides is a particular way of doing this, which is designed to be reasonably efficient for implementing and experimenting with rewrite rules corresponding to AI algorithms like reasoning engines, evolutionary learning processes, algorithmic chemistry, neural architecture learning and so forth, such as have been explored in the context of the OpenCog project.

\subsection{Formalizing MeTTa Structures and Spaces}

A bit more formally, construction of MeTTa in the metagraph context starts with:

\begin{itemize}
\item A space $S$ of symbols, which can be taken as labels of metagraph edges or targets
\item A grounding map $G: S \rightarrow F$, which maps symbols into a grounding domain $F$.  This map may be represented in a metagraph by directed edges labeled ''G'' for grounding
\item A ''variable'' property which identifies certain symbols as variable-names (this could be implemented in many ways in a metagraph, e.g. via associating such symbols with a unary edge labeled ''V'')
\item A space $T$ of ''types,'' drawn from an appropriate (possibly dependent and/or probabilistic) type system.   Types are a subset of the possible edge labels.  Type-space comes with a notion of inheritance, according to which it can (hopefully tractably) be determined whether type $T_1$ is a sub-case of type $T_2$ and whether $x$ inhabits type $T_1$, etc.   Indication that a label is a type could be implemented formally e.g. by associating the label with a unary edge labeled ''T''
\item MeTTa type-assignment and  type-restriction symbols become labels on metagraph edges, where they can be found and utilized via the MeTTa interpretation process
\item An equality operator $A=B$ that has the rough interpretation: The interpretation process can replace $A$ with$ B$ (sometimes, depending on situation).  But from a syntactic perspective ''='' is just another metagraph edge label.
\end{itemize}

\subsubsection{Formalizing the Grounding Domain}

Now what does an element of the grounding domain $F$ look like?  In general, we may say it looks like a function from $E \cup F$ into $E \cup F$ --- where $E$ is the space of  MeTTa expressions (to be defined below).   

MeTTa expressions may refer to the labels of metagraph edges as constant "data", so we can handle groundings that involve e.g. strings or numerical tensors by considering these as edge labels.

This general approach encompasses both standard grounding functions involving strings and vectors and such, and also further potentials.  In general,  this approach allows us to model $F$ as a space of non-well-founded sets \cite{Aczel1988}.   These can be represented as "apgs", i.e. as labeled directed graphs, which are a species of metagraph.

\subsubsection{Symbolizing the Activation of a Sub-Metagraph Encoding a MeTTA Program}

Next we need to formalize the interpretation of a sub-metagraph as a MeTTa program to be enacted on a metagraph.   Let the label $@$ denote the ''application'' symbol, with the semantics that: When the MeTTa interpreter encounters an edge labeled with a grounding function symbol, which has attached to it a unary edge labeled with ''@'', then it applies the grounding function to its specified arguments.

When a grounding function $f$ is applied to an expression $e$, then if $e$ is a grounding function with $@$ attached, $f$ first executes e and then acts on the result.  Otherwise, $f$ assumes $e$ is a ''literal expression'', and acts on the metagraph $e$ directly.

For the formalization of MeTTa as metagraph rewriting to work out in full glory, one would like the grounding functions to be implemented as metagraph rewriting rules also -- so the grounding functions should in principle be metagraphs represented via "MeTTa expressions" headed by application symbols.   This means that the matching of a grounding function to a metagraph argument should follow the "pattern match as metagraph homomorphism" approach described above, which seems sensible and unproblematic.

In actual practice, however, grounding functions in MeTTa programs will often be functions written in other programming languages (e.g. Python, Julia, Rust) to which MeTTa is connected.   This is not problematic in practice and often will be in the interest of greater usability and efficiency, but it's not a formal or semantic necessity.   So long as it works out equivalent (except for efficiency and UX wise) to what would happen with a pure MeTTa implementation, it's all good as regards the formal analyses and models given here.

\subsection{Formalizing the MeTTa Interpretation Process}

To complete the MeTTa picture we then need to introduce a few further processes apart from basic execution as encoded by @:

\begin{itemize}
\item An equality query "= x \$t" whose interpretation tries to find something to fill in the variable slot $\$t$ fulfilling the given equality.   Clearly the filling in of the slot is a species of metagraph rewriting.  
\item  A VBTO (variable binding term operator  \cite{VBTO} ) that takes an expression and binds one of its variables to some symbol or expression.  It must obey type inheritance restrictions.   Clearly VBTO is a species of metagraph rewriting.
\item A Pattern Matcher (PM), which when given an expression as a query tries to find expressions on which you can do VBTO for all the variables in the expression in a consistent way.
\item A "Transform" query type, which executes a PM query and then returns a version of the query that is transformed via having its variables filled-in with appropriate concrete values
\end{itemize}

A MeTTA {\bf expression} is then defined as one of the four cases: a grounded function application headed by $@$, a Pattern Matcher  or Transform query or an equality query.  

The interpretation process begins with an expression, whose interpretation then leads to other expressions being passed to the interpreter for interpretation, etc.  The action of the interpreter, and its intermediate states, can be represented as metagraphs, which is standardly done inside functional programming interpreters.  The interpreter is then carrying out a series of metagraph rewrites.   

Among the various processes involved here: An expression with variables can basically be understood as function whose interpretation requires the PM, i.e. it's a boolean predicate on the space of expressions.  So the MeTTa interpreter internally requires an EVB (expression variable binding) function $B$ that (repeatedly using VBTO) maps expressions into boolean predicates on expression-space, and the PM is the back end of $B$.

One can fully formalize the interpretation process using appropriate interpreter-focused metagraph edge labels such as ''Evaluate Next''.  The interpreter's process then evaluates the edges labeled ''Evaluate Next'' , which rewrite the metagraph accordingly, and part of this rewriting is adding ''Evaluate Next'' to some other edges, etc.

The formulation of MeTTa interpretation as execution of metagraph rewrite rules relies on the fact that the four processes listed above can be implemented as metagraph rewrite rules, i.e. as metagraph-level pattern matches formulated in terms of metagraph homomorphisms, followed by metagraph transformations (that may be non-surjective and non-total, potentially involving e.g. inserting new edges into the metagraph).   This implementability basically boils down to the implementation of the pattern matching underlying VBTO, potentially with consistency or equality constraints, as a metagraph homomorphism.   But this is evident.   The substitution of a variable with a referent in a metagraph $M$ does not modify the structure of $M$, and the nature of this matching is that the type of the referent must inherit from the specified type constraint on the variable, which comes along from the definition of homomorphism given above.  

\section{Rediscovering the Ruliad: From Metagraph Rewrite Rule Systems to the ($\infty$,1)-topos}

One of the consequences of the MeTTa framework is that metagraph rewrite rules, specified as MeTTa programs, can be expressed as metagraph content.   The execution trace of a MeTTa program can also be expressed as metagraph content.   Execution of a MeTTa program is then a process of "activating" a sub-metagraph, which then results in the enaction of the series of metagraph rewrite rules encoded by that sub-metagraph, and recording of the execution trace of the series of rewrites in an auxiliary "history" portion of the metagraph.   Proofs as studied in Homotopy Type Theory, including identity proofs, are among these execution traces.

This makes it automatic to create MeTTa programs whose inputs and outputs are MeTTA programs.   One can extrapolate to n'th order MeTTa programs that transform or create MeTTa programs that transform or create MeTTa programs, etc.    It is also straightforward to create MeTTa programs that modify themselves, effectively "infinite order" transformations.

Putting these aspects together, the collection of arbitrary-order MeTTa programs modifying arbitrary metagraphs becomes equivalent to what the Wolfram Physics community has come to call the "Ruliad" \cite{Ruliad}\cite{Ruliad-Wolfram} -- the pool of formal rules that modify formal rules that modify formal rules ... infinitely nested.    In \cite{Ruliad} it is shown that the Ruliad can be modeled as the $(\infty, 1)$-topos, which is basically an $\infty$-category defined over an $\infty$-groupoid.   The population of MeTTa programs operating on arbitrary metagraphs is a different formulation of the Ruliad, because the Wolfram community is concerned with hypergraphs rather than metagraphs, but there is a bisimulation between metagraph rewrite rules and hypergraph rewrite rules, which means that the two ultimately come down to the same thing (unless one is concerned with computational efficiency issues).

The practical value of this theoretical perspective is not yet clear.  Working computational efficiency issues into this Ruliad type framework involves numerous subtleties which are mostly currently unresolved and are critical in any practical applications.   However, it is nice to know what sort of formal structure one is dealing with, and one expects the theory and practice of MeTTa will evolve together over the coming years.

\section{Toward Efficient Algorithms for Practical Metagraph Pattern Matching}

Efficient interpretation of MeTTa programs is an unexplored domain which has multiple aspects fundamentally different from those encountered in other programming languages.   The key resource-intensive operations are pattern matching, and inference on equality queries.   Inference operationally reduces to pattern matching queries plus simple substitutions; so if pattern matching is made efficient, then managing the cost of inference is essentially down to inference control.

Both inference control and pattern matching are in essence metagraph search operations, which can be formulated via chronomorphisms and other metagraph fold operations as outlined in \cite{GTGI} \cite{goertzel2021patterns}.   However, folding over an entire large metagraph will often be undesirable due to resource issues, so it will often be desirable to prioritize which edges to fold over first, and then sometimes to halt a fold mid-way after it's dealt with higher-priority edges.   The key then becomes the dynamic prioritization process, which must be driven by two approaches:

\begin{itemize}
\item History-based pattern recognition, which is related to reinforcement learning and can be formalized as a discrete decision process integrating various AI techniques \cite{GTGI}
\item Heuristics chosen based on specific characteristics of the metagraph in question, e.g. the nature of the types and other enrichments and the statistics of edge sizes
\end{itemize}

If the metagraph has a distributed implementation this must be taken into account here as well, in that some of the search operations will involve finding things that are on remote machines relative to the search process, or are in persistent storage.   The PM functions that are usable against remotely or persistently stored knowledge may be an improper subset of those usable against the metagraph portion stored in RAM locally to the search process.   The speed constraints and formal restrictions of distributed PM must be taken into account in both history-based and heuristic guidance of the PM and equality inference search processes.

In \cite{GrGen} it is shown that guidance of the search process underlying graph pattern matching can be reasonably efficiently implemented using the framework of (SPO) algebraic graph rewriting.   In \cite{verigraph} it is shown that categorial formalization of graph rewriting can be quite directly use to structure the optimization of graph rewriting to account for particular graph characteristics.   It seems a similar strategy can likely be taken with the search processes underlying metagraph pattern matching and equality inference.     In essence one formulates heuristics guiding search in terms of predicates that act on sub-metagraphs, assessing their priority for search based on the context.   Such predicates can be learned based on history mining or formulated by human developers based on their knowledge of the particular metagraph in question.

\section{Conclusion}

This is an interim document summarizing ideas and suggested direction regarding an AGI R\&D project that is currently in process.   Achieving AGI is almost surely going to involve a subtle combination of theoretical and practical work, and the present paper embodies this combination in multiple ways.   

The design of the MeTTa language represents a synthesis of ideas from type theory, intuitionistic logic and other theoretical domains with practical experience gained via decades of experimentation with OpenCog Classic and earlier similar software systems.   Prior work expressing various forms of knowledge and cognitive process in MeTTa precursor languages and formalisms suggests that MeTTa may be capable to serve as a "language of thought" spanning the multiple knowledge and dynamics types characterizing human-level intelligence as well as the leading paradigms in the AI field.   Like this prior work, the process of fleshing out MeTTa is also being undertaken as a combination of theory and practice.  Following the completion of the first MeTTa prototype, there will be a phase of experimental implementation of various OpenCog-related AI algorithms in MeTTa, which will guide the practicalities of further development of the language front and back ends in various ways.    

The motivation for the present paper is not just the natural way that a formalization of MeTTa in terms of metagraph rewrite rules fits into the overall OpenCog Hyperon conceptual picture, but also the intuition that such a formalization may be helpful for the process of building and utilizing Hyperon in multiple respects, including guiding aspects of the back-end MeTTa interpreter development and the MeTTa implementation of various AI methods.   Functional programming interpreter implementation is often centered on graph rewriting, and many of the OpenCog-related AI methods are founded on mathematics that connects directly to metagraph representations and rewrite rules, so the metagraph-rewrite-rule formalization of MeTTa seems to have many "cognitive synergies" with other threads of Hyperon formalization, implementation and experimentation.

 \bibliographystyle{alpha}




% LNCS Homepage, \url{http://www.springer.com/lncs}. Last accessed 4
% Oct 2017ß
% \end{thebibliography}
\bibliography{metta-graphs.bib}

\end{document}